\documentclass[%
reprint,
superscriptaddress,
amsmath,amssymb,
prl,
floatfix,
]{revtex4-2}

\usepackage{float}
\usepackage{graphicx}
\usepackage{dcolumn}
\usepackage{bm}
\usepackage[colorlinks,linkcolor=blue,anchorcolor=blue,citecolor=blue,urlcolor=blue]{hyperref}
\usepackage{longtable}

\begin{document}

\title{Arbitrary Non-equilibrium Steady State Construction with a Levitated Nanoparticle}
\author{Yu Zheng}
\author{Lyu-Hang Liu}
\affiliation{CAS Key Lab of Quantum Information, University of Science and Technology of China, Hefei 230026, China}
\affiliation{CAS Center For Excellence in Quantum Information and Quantum Physics, University of Science and Technology of China, Hefei 230026, China}
\author{Xiang-Dong Chen}
\author{Guang-Can Guo}
\author{Fang-Wen Sun}
\email{fwsun@ustc.edu.cn}
\affiliation{CAS Key Lab of Quantum Information, University of Science and Technology of China, Hefei 230026, China}
\affiliation{CAS Center For Excellence in Quantum Information and Quantum Physics, University of Science and Technology of China, Hefei 230026, China}
\affiliation{Hefei National Laboratory, University of Science and Technology of China, Hefei 230088, China}

\date{\today }
\begin{abstract}
Non-equilibrium thermodynamics provides a general framework for understanding non-equilibrium processes, particularly in small systems that are typically far from equilibrium and dominated by fluctuations. However, the experimental investigation of non-equilibrium thermodynamics remains challenging due to the lack of approaches to precisely manipulate non-equilibrium states and dynamics. Here, \textcolor{black}{by shaping the effective potential of energy}, we propose a general method to construct a non-equilibrium steady state (NESS) with arbitrary energy distribution. Using a well-designed energy-dependent feedback damping, the dynamics of an optically levitated nanoparticle in vacuum is manipulated and driven into a NESS with the desired energy distribution. Based on this approach, a phonon laser state is constructed with an ultra-narrow linewidth of $6.40\text{ }\mu\text{Hz}$. Such an arbitrary NESS construction method provides a new approach to manipulating the dynamics processes of micromechanical systems and paves the way for the systematic study of non-equilibrium dynamics in interdisciplinary research fields. 
\end{abstract}

\maketitle

\begin{figure}[t]
	\includegraphics[width=0.5\textwidth]{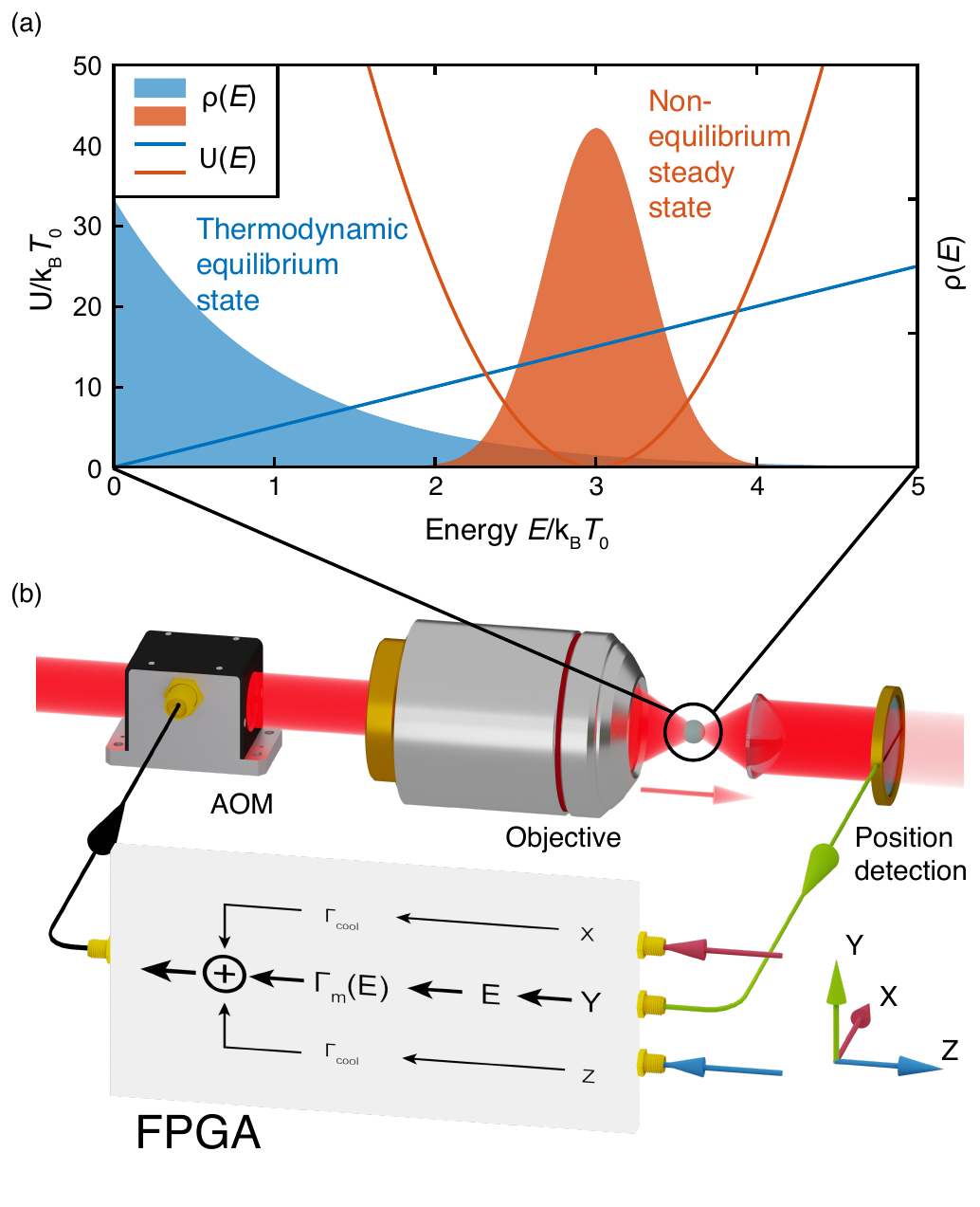}
	\caption{Schematic diagram of the construction of arbitrary NESSs. (a)  Modification of the \textcolor{black}{energy effective potential} $U(E)$ (solid lines) will change the corresponding energy distribution $\rho(E)$ (colored areas). \textcolor{black}{(b)} Experimental configuration. The energy distribution of a silica nanoparticle (radius $\sim$ 75 nm) trapped by a tightly focused laser beam is modified by the feedback control damping ($\Gamma_\text{m}(E)$), which is based on the real-time measurement of the translational degrees of freedom of the nanoparticle.}
	\label{fig:1}
\end{figure}

Originating from Maxwell's demon, a heat engine with feedback can break the second law of thermodynamics with the help of its microscopic state information\cite{Parrondo2015,Lutz2015}. 
Since the system can be controllably pushed away from equilibrium, it is ideally suitable for studying non-equilibrium dynamics. This is of importance not only in physics but also in the life and chemical sciences, where fluctuating systems far from equilibrium are a more common circumstance\cite{seifert2012stochastic,gnesotto2018broken,Amano2022}.
With extraordinary abilities to track and manipulate the dynamics of micro- and nano-particle, optical tweezers have become a standard experimental platform for microscopic thermodynamic research. More recently, optical tweezers and levitation in vacuum have shown excellent performance in demonstrations of fundamental physics\cite{Rondin2017,Rashid2016,Militaru2021}, macroscopic quantum mechanics\cite{Aspelmeyer2020,Tebbenjohanns2021,Magrini2021}, precision measurements\cite{ahn2020,ricci2019,Monteiro2017,Hempston2017,Rodenburg2016}, and in particular microscopic thermodynamics\cite{Li2010,Gieseler2014,Gieseler2015,Debiossac2020,Militaru2021_Kovacs,Gieseler2018,sheng2022nonequilibrium}. The ability to create an NESS and manipulate the strength of environmental interactions makes it appropriate for detailed studies of non-equilibrium thermodynamics under the influence of fluctuations\cite{Rondin2017,Militaru2021_Kovacs}. 

However, existing non-equilibrium experimental preparations are `scheme-to-state' approaches that rely on particular feedback control schemes to generate specific NESSs that correspond to the schemes\cite{Pettit2019,Gieseler2014}. A general design principle starting from any desired state remains to be investigated. Here, we introduce a universal approach based on \textcolor{black}{the shaping of the energy effective potential\cite{Gieseler2018}} that allows the construction of an arbitrary NESS with the help of energy-dependent feedback damping. A variety of NESSs, including phonon laser state, can be constructed using this approach. These customized motion states can be used for the investigation of non-equilibrium thermodynamics and precision measurements. Moreover, this prototype scheme can be further developed for manipulating levitated macroscopic quantum states\cite{Aspelmeyer2020,Tebbenjohanns2021,Magrini2021}.

Here, we consider an optically levitated nanoparticle in vacuum with an air damping $\Gamma_0$. Without any external interaction, the steady state of the nanoparticle will be a thermodynamic equilibrium state.
To obtain an NESS, extra channels for the exchange of energy or material are necessary. A damping rate $\Gamma_\text{m}$ is used to describe the rate and direction of energy exchange.
Here, we deploy an energy-dependent damping $\Gamma_\text{m}(E)$ to try to drive the system into an NESS.
In this case, the particle's energy dynamics can be manifested as a \textcolor{black}{Markovian stochastic process}.
Its dynamics are similar to an overdamped Brownian motion. The stochastic dynamic of a levitated nanoparticle's mechanical energy $E$ can be described with a Langevin equation\cite{Gieseler2014} (also see Supplemental Material (SM) for more details\cite{SM}), and we can \textcolor{black}{obtain the energy effective potential}, which is
\begin{equation}
	U(E)=\frac{1}{\Gamma_0}\int [\Gamma_\text{m}(E)+\Gamma_0]\mathrm{d}E
	\text{.}  \label{potential}
\end{equation}

Therefore, the distribution of $E$ corresponding to Eq. (\ref{potential}) can be given as a Boltzmann distribution\textcolor{black}{\cite{Gieseler2014,Gieseler2018}},
\begin{equation}
	\rho(E) =\frac{1}{Z} \exp \left[-\beta_0 U(E)\right]
	\text{,}
	\label{density}
\end{equation}
where $Z=\int_0^\infty \exp \left[-\beta_0 U(E)\right]\mathrm{d}E$, $\beta_0=1/k_\text{B}T_0$, $k_\text{B}$ is the Boltzmann constant, and $T_0$ is the particle's center of mass motion temperature under thermodynamic equilibrium.

From Eq. (\ref{density}), we are able to manipulate the feedback damping as
\begin{equation}
\Gamma_\text{m}(E)=-\frac{\Gamma_{0}}{\beta_{0}} \frac{1}{\rho(E)} \frac{d \rho(E)}{d E}-\Gamma_{0}
\text{,}
\label{dampE}
\end{equation}
and create a specific NESS with energy distribution $\rho(E)$ by deploying this $\Gamma_\text{m}(E)$ to the system.

In the experiment, we verify the feasibility of the construction of an arbitrary NESS of a levitated nanoparticle. As shown in Fig. \ref{fig:1}(b), a silica nanosphere with a diameter of approximately $150$ nm is trapped in vacuum by an optical potential with a tightly focused linearly polarized $1064$ nm laser.
We monitor the particle's real-time position and obtain its energy $E$ with a custom programmed field programmable gate array (FPGA) board.
The energy-dependent damping $\Gamma_\text{m}(E)$ can be added to the system by modulating the trapping laser power through the parametric feedback control protocol\cite{Gieseler2012,Zheng2020,Zheng2019}.
By controlling the depth and phase of the parametric feedback control signal, it is able to generate the energy-dependent feedback damping rate in an achievable range\cite{Zheng2020,Zheng2019}.
Therefore, it is possible to obtain the desired $\rho(E)$ by deploying the designed $\Gamma_\text{m}(E)$.

\begin{figure*}[t]
	\includegraphics[width=1\textwidth]{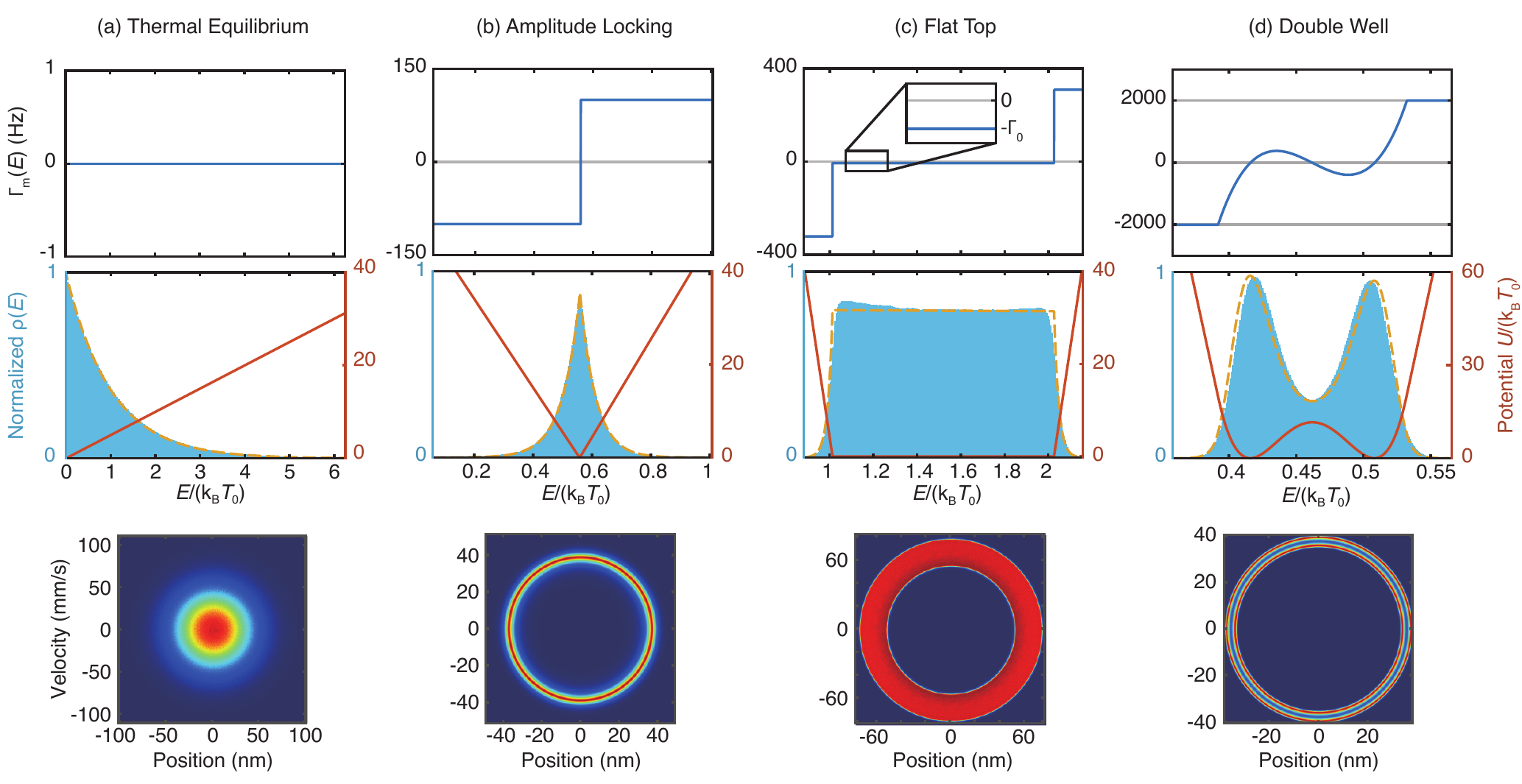}
	\caption{Experiment result of NESS construction under different $\Gamma_\text{m}(E)$. (a) Thermal equilibrium state as a comparison. (b)-(d) Three types of NESS constructions result, which is amplitude locking state by a step function $\Gamma_\text{m}(E)$, flat-top distributed state, and double well state. (Top) $\Gamma_\text{m}(E)$ deployed for the construction of each state. (Middle) Energy \textcolor{black}{effective} potential $U(E)$ and the measurement energy distribution $\rho(E)$ under $\Gamma_\text{m}(E)$ from each state. The solid lines are $U(E)$ according to Eq. (\ref{potential}). The dashed lines are the theoretical expectations of the energy distribution according to Eq. (\ref{density}). (Bottom) Phase plots of the measured oscillator's motion from each state. The air pressure is $10^{-3}$ mbar during the data collection. \textcolor{black}{The recording duration is 500 s for (a), (c) and 50 s for (b), (d).}}

	\label{fig:2}
	\newpage
	
\end{figure*}

Figure \ref{fig:2} shows the NESS construction results with three different $\Gamma_\text{m}(E)$. Moreover, a thermal equilibrium state with $\Gamma_\text{m}=0$ is shown in Fig. \ref{fig:2}(a) as a comparison.

As shown in Fig. \ref{fig:2}(b), $\Gamma_\text{m}(E)$ with a step function can be used to lock the oscillation amplitude of the levitated nanoparticle, which has been applied in a high-accuracy position and
mass measurement\cite{Zheng2020}. When the energy of the oscillator is lower (higher) than the target energy, a fixed negative (positive) feedback damping is applied to increase (decrease) the energy of the oscillator.
Such a two-stage step function creates a V-type $U(E)$, corresponding to a wedge shape $\rho(E)$.

We can construct an interesting NESS with a flat-top energy distribution, which can be used in the simulation of a free Brownian particle's diffusion process.

From Eq. (\ref{dampE}), a continuous uniform distribution of energy, which means $ d \rho(E)/d E=0$, requires $\Gamma_\text{m}(E)=-\Gamma_0$.
In other words, feedback damping is required to accurately offset the air damping to create the flat top.
To fulfill the requirement, a $-\Gamma_0$ part is inserted into a step function $\Gamma_\text{m}(E)$, as shown in Fig. \ref{fig:2}(c).
It can be observed that the oscillator's energy distribution is almost uniform in the $-\Gamma_0$ part. The slight fluctuation is caused by the vacuum pressure drift during data collection.

Finally, we attempt to make a double-well potential in energy, which is significant in bistable state studies such as Kramers turnover\cite{Rondin2017} or Landauer's principle\cite{Berut2012}.
Similar to the potential well structure in space, according to Eq. (\ref{potential}), it is feasible to construct a double-well $U(E)$ with a cubic function $\Gamma_\text{m}(E)$, as shown in Fig. \ref{fig:2}(d).
Because the maximum achievable feedback damping rate in our system is $\pm$2000 Hz, parts of $\Gamma_\text{m}(E)$ that exceed the limitation are truncated.
The experimental result shows that the oscillator has a twin-peak energy distribution, and its phase plot has a double-ring pattern.
Incidentally, the cubic function $\Gamma_\text{m}(E)$ used in double-well potential construction is compensated with a $-\Gamma_0$. Otherwise, the energy distribution will be asymmetric.

Moreover, the phonon laser is one of the most important NESS states, which can be utilized as a coherent phonon source or as an ultra-sensitive sensor\cite{Pettit2019,Vahala2009,Grudinin2009,Zhang2018,liu2021phonon}. 
Based on this NESS construction platform, we can concisely create a phonon laser state by a well-designed $U(E)$, which corresponds to $U(N)$ with $N=E/\hbar \Omega_0$, where $N$ is the phonon number and $\Omega_0$ is the eigenfrequency of the oscillator.
The phonon number distribution of the phonon laser that is well above the threshold will show a Gaussian distribution, which corresponds to a quadratic $U(N)$.
Therefore, according to Eq. (\ref{potential}), a phonon laser can be constructed by deploying a linear function $\Gamma_\text{m}(N)$ to the nano-oscillator\cite{Pettit2019}, that is

\begin{equation}
	\Gamma_\text{m}(N)=\gamma_{c} N-\gamma_{a}
	\text{,}
	\label{PLaser}
\end{equation}
where $\gamma_{a}$ is the linear gain factor and $\gamma_{c}$ is the nonlinear cooling factor. The dynamical equation of the phonon number can be written as\cite{SM}
\begin{equation}
\dot{N}=\left(\gamma_{a}-\Gamma_{0}\right) N-\gamma_{c} N^{2}+\frac{ \Gamma_0 k_{\mathrm{B}} T_{0}}{\hbar\Omega_0}+A
\text{,}
\label{PLDE}
\end{equation}
where $A=\sqrt{{2N\Gamma_0 k_{\mathrm{B}} T_{0}}/{\hbar\Omega_0}}{\mathrm{d} {W}}/{\mathrm{d} t}$ is the stochastic part and ${W}$ is the Wiener process.

According to Eq. (\ref{density}), the phonon number distribution fulfills
\begin{equation}
	\rho(N) =\frac{1}{Z_N} \exp \left\{-\beta_{0}\left(\frac{\hbar \Omega_{0} \gamma_{c}}{2 \Gamma_{0}}\left[N-\frac{\left(\gamma_{a}-\Gamma_{0}\right)}{\gamma_{c}}\right]^{2}\right)\right\}
	\text{,}
	\label{PLdensity}
\end{equation}
where $Z_N$ is the normalization factor. Eq. (\ref{PLdensity}) indicates that $\rho(N)$ under the driven of $\Gamma_\text{m}(N)$ is a Gaussian distribution with only the positive half axis part.

\begin{figure}[t]
	\includegraphics[width=0.44\textwidth]{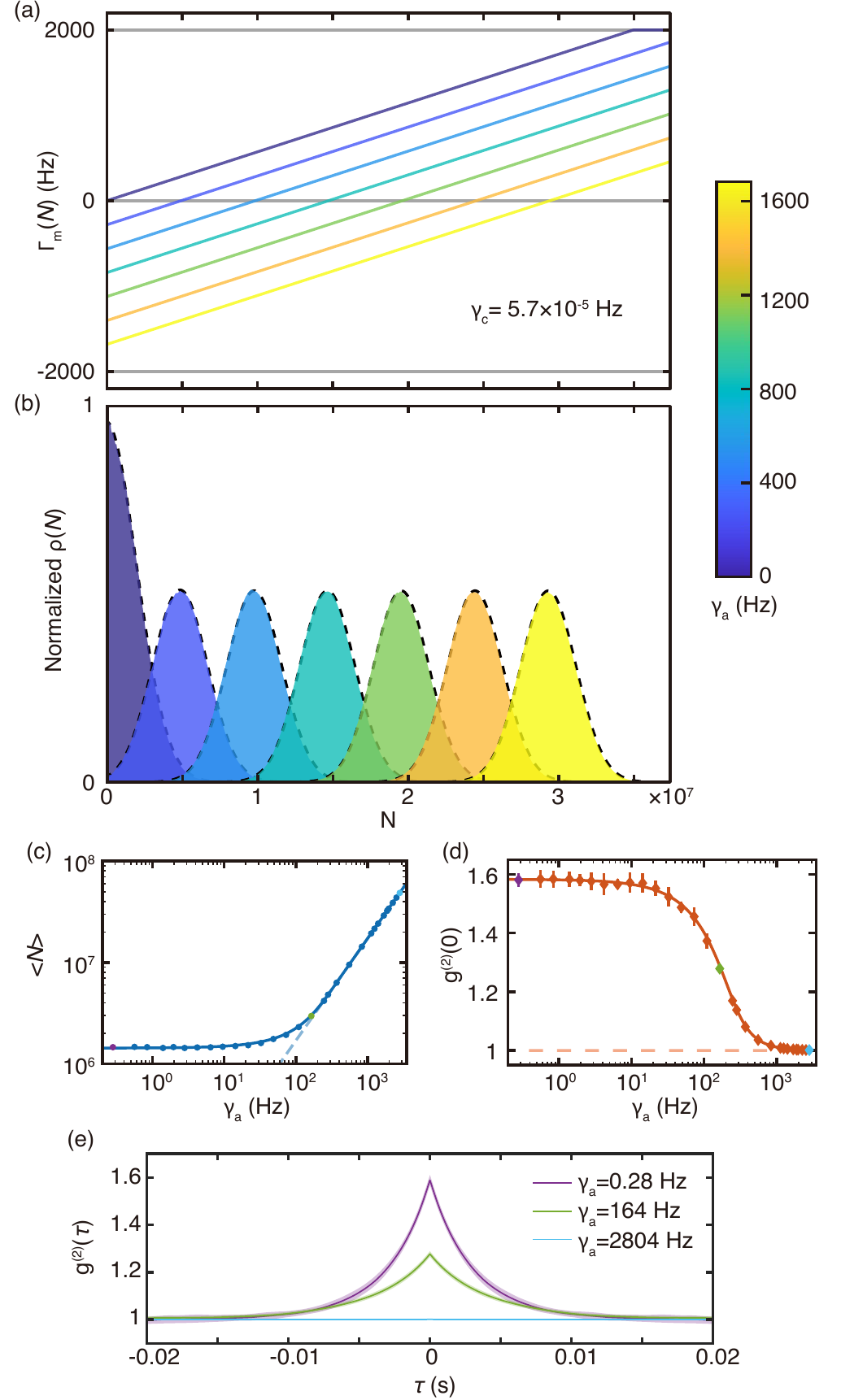}
	\caption{Experiment result of phonon laser construction. (a) The phonon-dependent feedback damping $\Gamma_\text{m}(N)$ with a fixed $\gamma_c$ and an increasing $\gamma_a$ that is deployed on the trapped nanoparticle. (b) The measured phonon number distribution of the nanoparticle driven by $\Gamma_\text{m}(N)$ from (a). The dashed lines are the theoretical expectations according to Eq. (\ref{PLdensity}). (c) Mean phonon number of the oscillator as a function of the gain factor $\gamma_a$. The dashed line shows $\langle N\rangle={\left(\gamma_{a}-\Gamma_{0}\right)}/{\gamma_{c}}$. The error bars are smaller than the data mark. (d) Second-order phonon autocorrelation function at zero delay, $g^{(2)}(0)$, as a function of the gain factor $\gamma_a$. The solid lines in (c) and (d) are theoretical expectations based on Eq. (\ref{PLdensity}). \textcolor{black}{(e) $g^{(2)}(\tau)$ with different $\gamma_a$. the selected point is marked with the same color in (c) and (d). The error is marked with the color areas.}
 The standard deviation represented by error bars or areas in (c), (d) and (e) is calculated from $10$ measurements. In these figures, $\gamma_c$ is a constant with $\gamma_c=5.7\times10^{-5}$ Hz, and the pressure is $10^{-3}$ mbar.
	}
	\label{fig:3}
 \newpage
\end{figure}

In the experiment, we construct different $\Gamma_\text{m}(N)$ to obtain phonon lasers with various phonon number distributions. As shown in Figs. \ref{fig:3}(a) and \ref{fig:3}(b), the mean phonon number of the oscillator is increased with the increasing $\gamma_{a}$. Meanwhile, as the $\gamma_{c}$ keeps constant during the experiment, the shape of the phonon number distribution $\rho(N)$ is remain the same.

To study the phonon laser properties of the nano-oscillator driven with $\Gamma_\text{m}(N)$, the linear gain factor $\gamma_a$ is selected as the pump power coefficient. The threshold property of a laser is verified by increasing $\gamma_a$ from 0 Hz. Figure \ref{fig:3}(c) shows that when $\gamma_a$ exceeds a threshold, the mean phonon number $\langle N\rangle$ increases linearly with $\gamma_a$, where $\langle N\rangle={\left(\gamma_{a}-\Gamma_{0}\right)}/{\gamma_{c}}$. \textcolor{black}{As shown in Fig. \ref{fig:3}(d,e), as $\gamma_a$ increases from zero to well above the threshold, $g^{(2)}(0)$ decreases to 1, which means that the oscillation changes from a thermal state to a coherent state.} It can be noticed that $g^{(2)}(0)$ does not start from $2$ when $\gamma_a=0$ Hz. This is because the nonlinear cooling factor $\gamma_c$ is a nonzero constant, which compels the system to deviate from a pure thermal state.

The narrowing of linewidth is another important feature of lasers\cite{Pettit2019,Grudinin2009}.
Utilizing the analysis of the stochastic phase noise, the full width at half maximum linewidth of a free-run phonon laser is supposed to be

	$\Delta f_\text{FWHM}={k_\text{B} T_{0} \Gamma_{0}}/{4 \pi \langle N\rangle \hbar \Omega_{0}}$\cite{SM}.
However, due to the nonlinearity of the optical potential, the dispersion of the phonon number would introduce a frequency shift that makes the linewidth much wider than the theoretical result.

\begin{figure}[t]
	\includegraphics[width=0.5\textwidth]{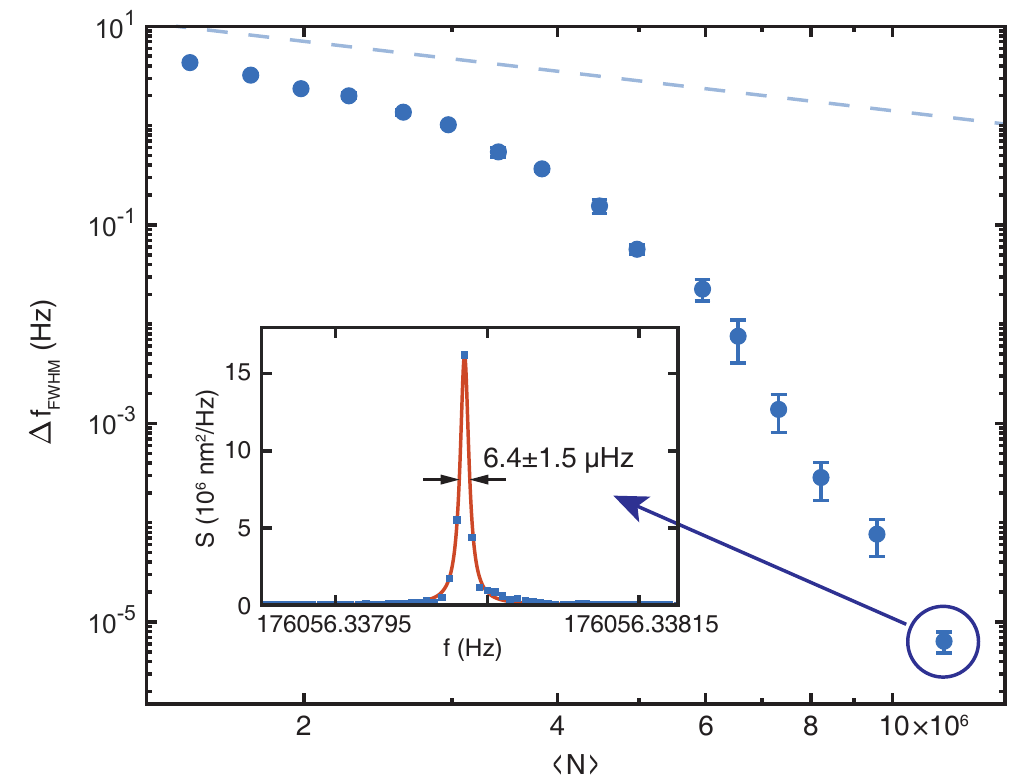}
	\caption{Power spectral density (PSD) linewidth $\Delta f_\text{FWHM}$ of phonon laser state as a function of the mean phonon number with feedback frequency stabilization. Error bars represent the standard deviations that are calculated from $5$ trajectories of each data point. The recording time of each sampling data point ranges from $200$ to $2\times10^5$ s depending on the linewidth required spectrum resolution. The dashed line is the theoretical linewidth of a free-run phonon laser. The inset figure is the averaged PSD of the selected data point trajectories. The solid line in the inset is a fitting of the Lorentzian function. The data are recorded at a pressure of $10^{-3}$ mbar. The corresponding phonon laser parameters are $\gamma_{c}=5\times10^{-5}$ Hz and $\gamma_{a}=20 \text{ to }600$ Hz.}
	\label{fig:4}
\end{figure}

To overcome this challenge, an active feedback frequency stabilization based on an integral feedback controller is deployed. The duration of each oscillation cycle is compared with the period corresponding to the locking frequency. The frequency error is compensated by modulating the base intensity of the trapping laser. As shown in Fig. \ref{fig:4}, under frequency stabilization, the linewidth of the phonon laser decreases when the mean phonon number increases. The measured phonon laser linewidth is much narrower than the theoretical free-run linewidth, which indicates that the phase noise error introduced by the stochastic and nonlinear process in the phonon laser is well suppressed by the frequency stabilization. The narrowest linewidth recorded in the experiment is $\Delta f_\text{FWHM}=6.40(\pm1.51) \text{ }\mu\text{Hz}$, and the corresponding quality factor is $\text{Q}=2.88(\pm0.71)\times10^{10}$. This is the highest quality factor of the translational oscillation of the levitated nanoparticle, which can be further applied in the precision measurement requiring long-term stabilization like ultra-weak gravity force detection\cite{westphal2021measurement}.

In conclusion, we have introduced an energy-dependent feedback damping to construct an NESS with an arbitrary energy distribution of an optically levitated nanoparticle. The feasibility of this method has been experimentally verified by demonstrating special steady states that have never been reported. Moreover, a phonon laser steady state with an ultra-narrow linewidth is produced by this method.
The energy flow control and state manipulation shown in this work could be used to reinforce the optical levitation as an excellent platform in the microscopic thermodynamic investigation. Furthermore, such a state construction method could be developed as a possible solution for levitated quantum state manipulation\textcolor{black}{, such as a phonon Fock state.}
The ability to manipulate a micro-system’s energy dynamic process explores a new approach to investigate fluctuating thermodynamics process, such as a Brownian motor\cite{Astumian1997,Bang2018} and Landauer's principle\cite{Berut2012}, in energy space with optical levitation. Finally, phonon lasers with high Q-factor can benefit the development of precision measurements based on levitated nano-senses\cite{Arita2020}, such as ultra-weak force detection.

We acknowledge support from  the National Natural Science Foundation of China (Grant Nos. 12104438, 62225506), CAS Project for Young Scientists in Basic Research (Grant No. YSBR-049), the Fundamental Research Funds for the Central Universities, and the Innovation Program
for Quantum Science and Technology 2021ZD0303200. The sample preparation was partially conducted at the USTC Center for Micro and Nanoscale Research and Fabrication.


%

\end{document}


\title{Supplemental Material: Arbitrary Non-equilibrium Steady State Construction with a Levitated Nanoparticle}
\author{Yu Zheng}
\author{Lyu-Hang Liu}
\affiliation{CAS Key Lab of Quantum Information, University of Science and Technology of China, Hefei 230026, China}
\affiliation{CAS Center For Excellence in Quantum Information and Quantum Physics, University of Science and Technology of China, Hefei 230026, China}
\author{Xiang-Dong Chen}
\author{Guang-Can Guo}
\author{Fang-Wen Sun}
\email{fwsun@ustc.edu.cn}
\affiliation{CAS Key Lab of Quantum Information, University of Science and Technology of China, Hefei 230026, China}
\affiliation{CAS Center For Excellence in Quantum Information and Quantum Physics, University of Science and Technology of China, Hefei 230026, China}
\affiliation{Hefei National Laboratory, University of Science and Technology of China, Hefei 230088, China}
	
	\maketitle
	
	\tableofcontents

	\makeatletter
	\renewcommand{\thefigure}{S\@arabic\c@figure}
	\makeatother
	\makeatletter
	\renewcommand{\thetable}{S\@arabic\c@table}
	\makeatother
	\makeatletter
	\renewcommand\thesection{\arabic{section}}
	\renewcommand\thesubsection{\thesection.\arabic{subsection}}
	\makeatother
	
\makeatletter
\renewcommand{\theequation}{S\@arabic\c@equation}
\makeatother
	
	\clearpage
	
	\section{Experiment Setup}
	
	\begin{figure}[t]
		\includegraphics[width=1\textwidth]{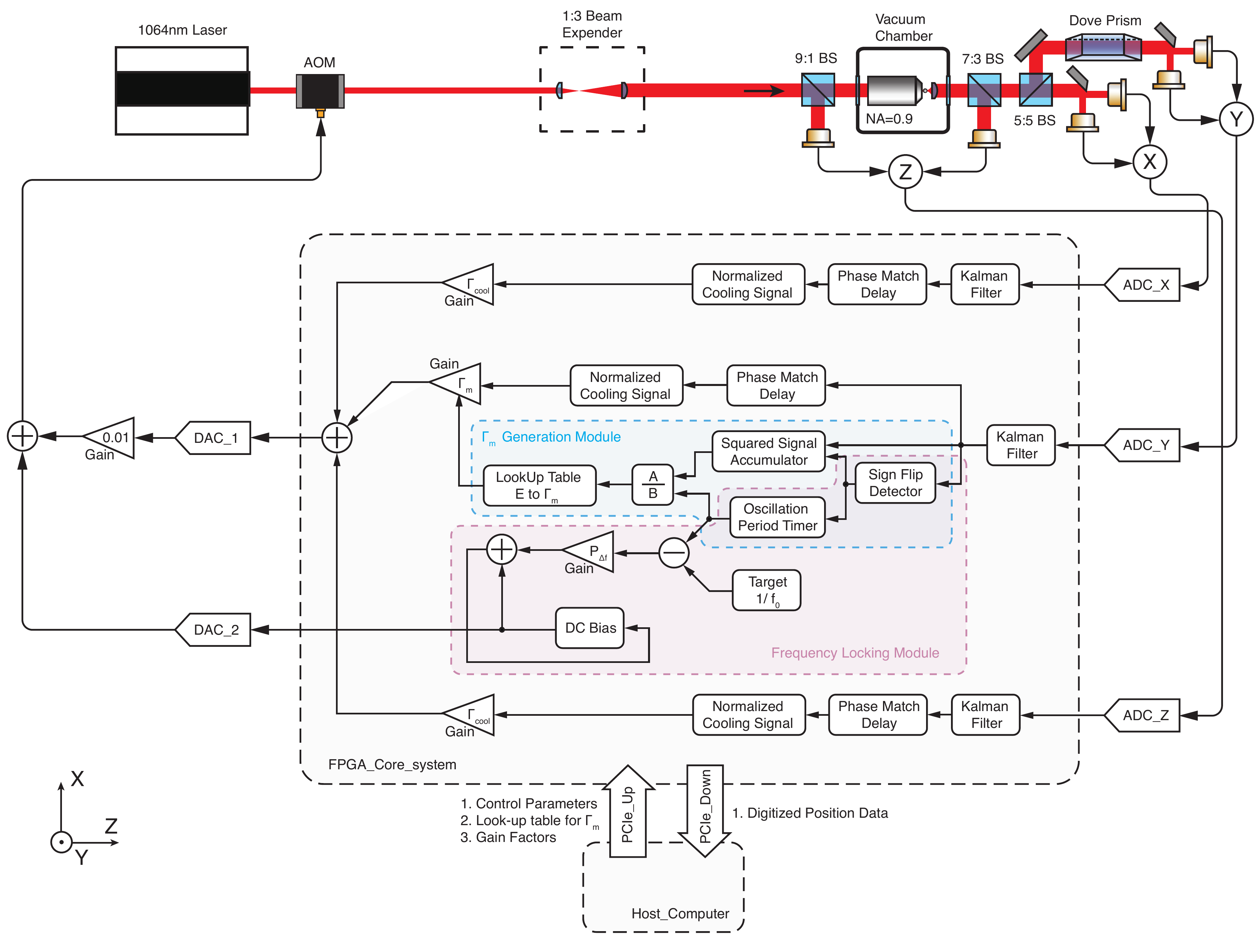}
		\caption{\textbf{Experimental configuration illustration.} }
		\label{figS1}
	\end{figure}
	
	This section describes the details of the experimental setup in the main text.
	The schematic of the experiment setup is shown in Fig. \ref{figS1}.
	
	\subsection{Device structure}
	A CW 1064 nm laser (Coherent Mephisto 2000) is used as the trapping laser. Its intensity is modulated by an acoustic-optic modulator (AOM). After the beam expand lens set, the laser beam with a diameter of approximately 4.5 mm is guided into an objective (N.A.=0.9, Nikon CFI LU Plan Fluor EPI 100X) inside a vacuum chamber. The laser intensity before the objective is measured to be 250 mW. The trapping laser is focused by the objective to form the optical potential for particle trapping. After the objective, an aspheric lens (N.A.=0.55, Thorlabs C230TME-1064) collects the forward scattering light and sends it to the particle position detection unit, which is consisted of three sets of homemade balanced photodetectors, to monitor the trapped particle's three motional degrees (set as X, Y, Z-axis) of freedom. A Dove prism is used to rotate the forward scattering light by $90^{\circ}$ for the convention of Y-axis motion detection. The position signals are sent to a field-programmable gate array (FPGA) board to generate the feedback signal.
	\subsection{FPGA program structure}
	The position voltage signals from balanced photodetectors are processed by an FMC card, which has 4 channels of 16-bit ADC and 2 channels of 14-bit DAC. The digitized position signals are transferred to the FPGA core (Xilinx Virtex UltraScale+ XCVU3P) to generate the feedback control signal.
	
	We focus on the nonequilibrium steady state (NESS) control signal generation of Y-axis motion. First, the position signal is processed by a Kalman filter to eliminate the out-band noise. Then, the signal is sent to three modules. One is used to generate a normalized cooling signal. One is used to calculate the energy of the oscillator for $\Gamma_\text{m}$ generation. And the last one is used to lock the oscillator's frequency.
	The details of cooling signal generation are described in Ref. \cite{Zheng2019}.
	
	Next, we discuss the process of $\Gamma_\text{m}$ generation.
	As the energy of a harmonic oscillator is proportional to the square of the amplitude, the energy calculation is accomplished by averaging the square of the position displacement of every data point in one oscillation cycle. A ``Sign Flip Detector" monitors the timing when the position signal's sign flips from negative to positive. When the sign flip event occurs, the ``Sign Flip Detector" sends a trigger signal to refresh the output data of the ``Squared Signal Accumulator" and the ``Oscillation Period Timer" and clear the two modules' counter. The output data of the two modules are divided to obtain the variable that is proportional to the energy.
	
	Then, the calculated energy $E$ is sent to a lookup table. The map of $\Gamma_\text{m}(E)$ is stored in the lookup table. We can obtain the feedback damping $\Gamma_\text{m}(E)$ that corresponds to the measured $E$. $\Gamma_\text{m}(E)$ multiplies the normalized cooling signal, and the parametric feedback control signal of the NESS is prepared finally.
	
	However, this energy calculation introduces an additional feedback delay time of one oscillation period. In most situations, the slow variation of the particle's energy in vacuum eliminates the effects of energy calculation delay. However, if the energy variation is too fast, the energy calculation delay's influence is observable. For example, the slight mismatch between the theoretical energy distribution and the measured distribution in Fig. 2h in the main text is because the feedback control cannot perfectly catch up with the switching events that the particle jumps between the two wells as the energy dramatically changes during well switching.
	
	The NESS control signal is added with the cooling signals of the X- and Z-axis and output through a DAC channel. The amplitude of the DAC output signal is reduced by 100 times with an amplifier to increase the control precision of the feedback damping. The output signal is added up with a DC bias voltage, which makes the AOM work in the linear regime and is sent to the RF driver of the AOM.
	
	The motion of the X- and Z-axis is cooled to about 10 K to minimize nonlinear coupling between different axes. It should be noted that the center of mass motion (COM) temperature of the above two axes should not be cooled too low; otherwise, the Y-axis signal with higher amplitude will sneak into the feedback cooling loop of the X- and Z-axis due to the cross-talk of position signals. This will introduce an unexpected cooling damping to the Y-axis motion, thereby undermining the accuracy of the feedback damping $\Gamma_\text{m}(E)$ applied to the Y-axis motion.
	
	The feedback frequency stabilization of a phonon laser is achieved by modulating the DC bias. As the oscillator's frequency is proportional to the square root of the laser intensity, the oscillator's frequency can be modulated by adjusting the DC bias of the AOM.
	
	A simple integral controller is utilized to lock the oscillation frequency. The output of the ``Oscillation Period Timer" is compared with the target period. The difference is multiplied by a gain factor and added to the current DC bias. The addition result is looped back to the DC bias variable and becomes the new DC bias. The refresh loop of the DC bias is also triggered by the ``Sign Flip Detector".
	
	The position signal processed in the FPGA is also transferred to a computer for data processing. Therefore, the ADC for data collection and the FPGA modules are driven by the same clock, which is necessary for ultra-narrow linewidth phonon laser experiments.

\begin{figure}[h]
		\includegraphics[width=1\textwidth]{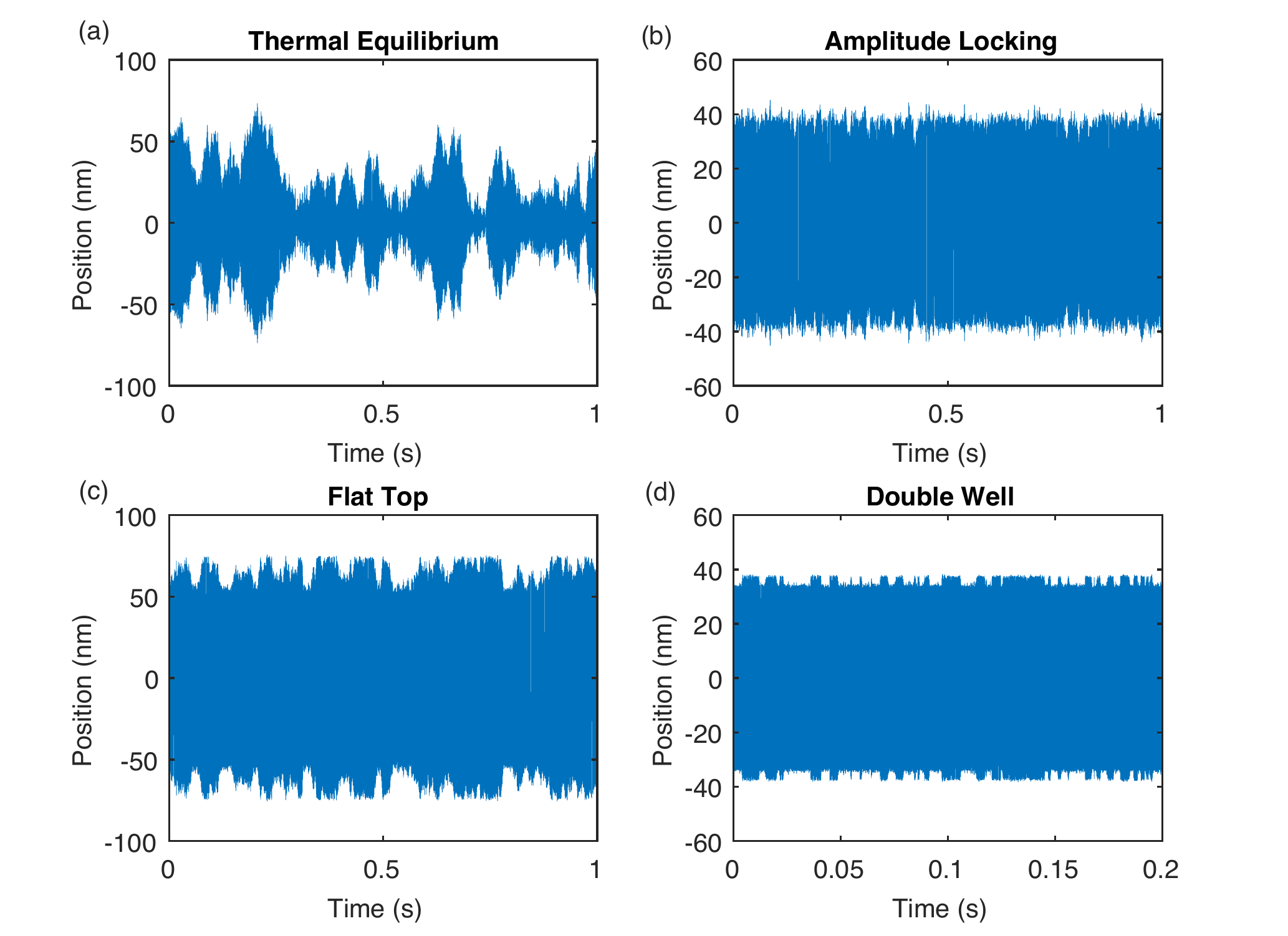}
		\caption{\textbf{Trajectories of the NESS.} (a) Thermal equilibrium state. (b) Amplitude locking state. (c) Flat-top state. (d) Double-well state.} 
		\label{Trajectory}
	\end{figure}

\textcolor{black}{\subsection{Trajectories of the NESS}
 The trajectories of the NESS in the Fig.2 of the main text is shown in the Fig.\ref{Trajectory}. The variation pattern of the oscillator amplitude can be seen in the envelope of the trajectories.}
 
 \textcolor{black}{\subsection{Data Processing for the Energy and Phase plot}}
The particle's position can be obtained from the calibrated photodetector signal. To obtain the particle's energy and phase plot, we also need to get the velocity of the particle. To obtain the velocity, we use the finite difference method. 

The detailed process is shown as follows. First, the raw position data is processed with a band-pass filter to eliminate out-band noise. And we have the position data $x_i$ with $i=1\sim N$, and the sampling interval of the position data is $\Delta t$.
Then, a cubic spline interpolation is used to double the sampling rate. So that $x'_{2i-1}=x_i$ is the measured position data and $x'_{2i}$ is the interpolated position data. And we have 
\begin{equation}
    v'_{j-1}=\frac{x'_{2j}-x'_{2(j-1)}}{\Delta t}\text{,}
\end{equation}
 with $j=2\sim N$. 

$v'_{j-1}$ is smaller than the true velocity due to the truncation error in the finite difference approximation. To compensate this error, $v'_{j-1}$ has to be multiplied by a factor $c$. For a sine function with frequency $f_0$, we have
\begin{equation}
    c=\frac{\pi f_0 \Delta t}{\sin(\pi f_0 \Delta t)}\text{.}
\end{equation}
The particle's velocity is $v_{j-1}=c\cdot v'_{j-1}$, where the particle's eigen-frequency is used to be $f_0$ for the calculation of $c$.

 The phase trajectory of the particle is $(x_{j}\text{, }v_{j-1})$. The phase plot is the distribution of the phase trajectory on the phase plane. The purpose of using interpolated data to calculate the velocity is to make the velocity samples match the phase of the position samples. Otherwise, the obtained phase plot would be skewed.

 The particle's energy can be calculated with 
 \begin{equation}
    E_{j-1}=\frac{1}{2}mv^2_{j-1}+\frac{1}{2}m\Omega_0^2 x_j^2\text{,}
\end{equation}
where $m$ is the mass of the particle, and $\Omega_0$ is the eigen angular frequency of the particle's oscillation.

	\section{Derivation of $U(E)$ and $\rho(E)$ under $\Gamma_\text{m}(E)$}
	This section shows the deviation of $\rho(E)$ and $U(E)$ in the main text from a levitated nanoparticle's equation of motion.
	The derivation process generally refers to the process in Ref. \cite{Gieseler2014}.
	
	Following the steps in Ref. \cite{Gieseler2014}, the motion of a trapped nanoparticle in an independent motion dimension is considered.
	The equation of motion can be written as
	\begin{equation}
		\ddot{q}(t)+\left[\Gamma_{0}+\Gamma_\text{m}(t)\right] \dot{q}(t)+\Omega_{0}^{2} q(t)=\frac{1}{m}F_{\text {random }}(t)
		\text{,}
		\label{EQM}
	\end{equation}
	where $q$ is the position of the particle, $\Gamma_{0}$ is the air damping rate, $\Gamma_\text{m}(E(t))$ is the feedback damping rate, $\Omega_{0}$ is the eigenfrequency of the particle's oscillation, $m$ is the particle's mass, $F_{\text {random }}(t)=\sqrt{2m\Gamma_0k_\text{B}T_0}\xi(t)$ is the stochastic force from the environment (air molecule collisions), $T_0$ is the COM temperature, and $\xi(t)=\mathrm{d} W(t) / \mathrm{d} t$, $W$ is the Wiener process.
	
	Rewriting equation (\ref{EQM}) into a stochastic differential equation (SDE), we have
	\begin{equation}
		\mathrm{d}q=\frac{p}{m}\mathrm{d}t
		\text{,}
		\label{SDEa}
	\end{equation}
	\begin{equation}
		\mathrm{d}p=\left[-m\Omega_0^2q-\Gamma_0p-\Gamma_\text{m}(t)p
		\right]\mathrm{d}t+\sqrt{2m\Gamma_0k_\text{B}T_0}\mathrm{d}W
		\text{,}
		\label{SDEb}
	\end{equation}
	where $p$ is the particle's momentum.
	
	Consider the dynamics in the particle's energy $E$. To avoid multiplicative noise, the dynamics of energy are described in $\epsilon$ such that $\epsilon=\sqrt{E}$. Neglecting all terms that higher the order of $(\mathrm{d}t)^{3/2}$, we have
	\begin{equation}
		\mathrm{d} \epsilon=\left(\frac{\partial \epsilon}{\partial q}\right) \mathrm{d} q+\left(\frac{\partial \epsilon}{\partial p}\right) \mathrm{d} p+\frac{1}{2}\left(\frac{\partial^{2} \epsilon}{\partial p^{2}}\right)(\mathrm{d} p)^{2}
		\text{.}
		\label{depsilon}
	\end{equation}
	With
	\begin{equation}
		E(q, p)=\frac{1}{2} m \Omega_{0}^{2} q^{2}+\frac{p^{2}}{2 m}
		\text{,}
		\label{EE}
	\end{equation}
	and equation (\ref{SDEa}) and (\ref{SDEb}), equation (\ref{depsilon}) can be obtained as
	
	\begin{equation}
		\begin{aligned}
			\mathrm{d} \epsilon&=m \Omega_{0}^{2} \frac{q}{2 \epsilon} \mathrm{d} q+\frac{1}{2 \epsilon} \frac{p}{m} \mathrm{d} p+\frac{1}{2}\left(\frac{1}{2 m \epsilon}-\frac{1}{4 \epsilon^{3}} \frac{p^{2}}{m^{2}}\right)(\mathrm{d} p)^{2}\\
			&=\frac{1}{2 \epsilon} \frac{p^2}{m} \left[-\Gamma_0+\Gamma_\text{m}(\epsilon)
			\right]\mathrm{d}t+\frac{1}{2 \epsilon} \frac{p}{m}\sqrt{2m\Gamma_0k_BT_0}\mathrm{d}W+\frac{1}{2\epsilon}\left(1-\frac{p^{2}}{2m \epsilon^{2}}\right){\Gamma_0k_BT_0}(\mathrm{d}W)^{2}
			\text{.}
			\label{de2}
		\end{aligned}
	\end{equation}
	
	Next, we use quasi-static approximations. The coherence time of the levitated oscillator in high vacuum is much longer than its oscillation period. In one period, the energy can be considered as a constant, and the velocity can be approximated as a sine function. We focus on the energy varying in one oscillation period, $	\tau=2 \pi / \Omega_{0}$, which is
	\begin{equation}
		\begin{aligned}
			\Delta \epsilon=&\int_{0}^{\tau} \mathrm{d} \epsilon\\
			=&-\frac{\Gamma_0}{2} \int_{0}^{\tau} \frac{p^{2}}{m \epsilon} \mathrm{d} t-\frac{1}{2} \int_{0}^{\tau} \frac{\Gamma_\text{m}(\epsilon)p^{2}}{m \epsilon} \mathrm{d} t\\
			&+\sqrt{2 m \Gamma_{0} k_{\mathrm{B}} T_{0}} \int_{0}^{\tau} \frac{p}{2 m \epsilon} \mathrm{d} W \\
			&+\Gamma_{0} k_{\mathrm{B}} T_{0} \int_{0}^{\tau} \frac{1}{2 \epsilon}\left(1-\frac{p^{2}}{2 m \epsilon^{2}}\right)(\mathrm{d} W)^{2}
			\text{,}
		\end{aligned}
	\end{equation}
	and we have\cite{Gieseler2014}
	\begin{equation}
		\Delta \epsilon=\left(-\frac{(\Gamma_{0}+\Gamma_\text{m}) \epsilon}{2}+\frac{\Gamma_{0} k_{\mathrm{B}} T_{0}}{4 \epsilon}\right) \tau+\sqrt{\frac{\Gamma_{0} k_{\mathrm{B}} T_{0}}{2}} W(\tau)
		\text{.}
		\label{De}
	\end{equation}
	Rewriting equation (\ref{De}) into a differentiated form, we have
	\begin{equation}
		\mathrm{d} \epsilon=\frac{1}{\nu} f(\epsilon) \mathrm{d} t+\sqrt{\frac{2 k_{\mathrm{B}} T_{0}}{\nu}} \mathrm{d} {W}
		\text{,}
		\label{Langevin}
	\end{equation}
	where $\nu=4 / \Gamma_{0}$ and
	\begin{equation}
		f(\epsilon)=-2\epsilon\left(\frac{\Gamma_0+\Gamma_\text{m}(\epsilon)}{\Gamma_0}\right)+\frac{ k_{\mathrm{B}} T_{0}}{ \epsilon}
		\text{.}
		\label{f}
	\end{equation}
	Equation (\ref{Langevin}) is a Langevin equation. We can obtain the energy properties of the nanoparticle from the general properties of a Langevin equation.
	
	The energy effective potential is
	\begin{equation}
		\begin{aligned}
			U_{\epsilon}(\epsilon)&=-\int f(\epsilon)\mathrm{d}\epsilon\\
			&=\epsilon^{2}+\frac{2}{\Gamma_0}\int\epsilon\Gamma_\text{m}(\epsilon)d\epsilon-k_{\mathrm{B}} T_{0} \ln \epsilon
			\text{.}
			\label{epotential}
		\end{aligned}
	\end{equation}
	The energy distribution follows the Maxwell–Boltzmann distribution, which is
	\begin{equation}
		\begin{aligned}
			\rho(\epsilon)&=\frac{1}{Z_\epsilon}\exp\left[-\beta_{0}U_{\epsilon}(\epsilon)\right]\\
			&=\frac{1}{Z_\epsilon}\epsilon \exp \left\{-\beta_{0}\left(\epsilon^{2}+\frac{2}{\Gamma_0}\int\epsilon\Gamma_\text{m}(\epsilon)d\epsilon\right)\right\}
			\text{,}
			\label{edensity}
		\end{aligned}
	\end{equation}
	where $\beta_0=1/(k_\text{B}T_0)$, $Z_\epsilon$ is the partition function.
	
	Change the variable in the distribution equation (\ref{edensity}) from $\epsilon$ to $E$. We have equation (2) in the main text, that is,
	\begin{equation}
		\rho(E) =\frac{1}{Z} \exp \left\{-\frac{\beta_{0}}{\Gamma_0}\int{[\Gamma_\text{m}(E)+\Gamma_0]}dE\right\}
		\text{.}
		\label{Edensity}
	\end{equation}
	Imitate the relation that $\rho(\epsilon)=\frac{1}{Z_\epsilon}\exp\left[-\beta_{0}U_{\epsilon}(\epsilon)\right]$.
	We can have the effective potential for the energy that is
	\begin{equation}
		U(E)=\frac{1}{\Gamma_0}\int [\Gamma_\text{m}(E)+\Gamma_0]\mathrm{d}E
		\text{,}  \label{potential}
	\end{equation}
	which is also equation (1) in the main text.
	
	We should note that $U(E)$ can be utilized to obtain the statistical properties of energy $E$, such as the distribution or mean value. However, the dynamic properties such as diffusion are better to be investigated using $\epsilon$ and $U_{\epsilon}(\epsilon)$, as the Langevin equation (\ref{Langevin}) can only be obtained with variable $\epsilon$.
	
	\section{Derivation of the phonon laser's dynamical equation}
	This section shows the deviation of the phonon laser's dynamical equation, which is equation (5) in the main text.
	
	Rewriting equation (\ref{Langevin}) into
	\begin{equation}
		\mathrm{d} \epsilon=\mu\mathrm{d} t+\sigma \mathrm{d}W
		\text{,}
		\label{itoe}
	\end{equation}
	where $\mu=-\epsilon\left[{\Gamma_0+\Gamma_\text{m}(\epsilon)}\right]/2+{ \Gamma_0 k_{\mathrm{B}} T_{0}}/{ 4\epsilon}$, and $\sigma=\sqrt{{\Gamma_0 k_{\mathrm{B}} T_{0}}/{2}}$.
	
	According to Itô's lemma and $E=\epsilon^2$, we have
	\begin{equation}
		\begin{aligned}
			\mathrm{d}E&=\left(\mu\frac{\partial E}{\partial\epsilon}+\frac{\sigma^2}{2}\frac{\partial^2 E}{\partial\epsilon^2}\right)\mathrm{d} t+\sigma\frac{\partial E}{\partial\epsilon} \mathrm{d} {W}\\
			&=(2\epsilon\mu+\sigma^2)\mathrm{d} t+2\sigma\epsilon\mathrm{d} {W}\\
			&=(-E\left[{\Gamma_0+\Gamma_\text{m}(E)}\right]+{ \Gamma_0 k_{\mathrm{B}} T_{0}})\mathrm{d} t+\sqrt{2E\Gamma_0 k_{\mathrm{B}} T_{0}}\mathrm{d} {W}
			\text{.}
			\label{dE}
		\end{aligned}
	\end{equation}
	Replace energy $E$ with phonon number $N\cdot\hbar\Omega_0$ and set the feedback damping $\Gamma_\text{m}$ to the phonon laser control damping, which is $\Gamma_\text{m}(N)=\gamma_{c} N-\gamma_{a}$. Equation (\ref{dE}) can be written as
	
	\begin{equation}
		\mathrm{d}N=(-N\left[{\Gamma_0+\Gamma_\text{m}(N)}\right]+\frac{ \Gamma_0 k_{\mathrm{B}} T_{0}}{\hbar\Omega_0})\mathrm{d} t+\sqrt{\frac{2N\Gamma_0 k_{\mathrm{B}} T_{0}}{\hbar\Omega_0}}\mathrm{d}{W}
		\text{.}
		\label{dN}
	\end{equation}
	
	We have the phonon laser's dynamical equation, which is
	\begin{equation}
		\dot{N}=\left(\gamma_{a}-\Gamma_{0}\right) N-\gamma_{c} N^{2}+\frac{ \Gamma_0 k_{\mathrm{B}} T_{0}}{\hbar\Omega_0}+\sqrt{\frac{2N\Gamma_0 k_{\mathrm{B}} T_{0}}{\hbar\Omega_0}}\frac{\mathrm{d}{W}}{\mathrm{d} t}
		\text{.}
		\label{phonondynamic}
	\end{equation}
	
	\section{Discussion of the phonon laser spectrum}
	\subsection{Derivation of a free-run phonon laser linewidth}
	For a coherent oscillator, its spectrum linewidth is mainly dependent on the oscillator's phase noise. To simplify the process to obtain the phonon laser's linewidth, we only consider the influence of the phase noise.
	
	For a harmonic oscillator at time $t_0$, its position and velocity can be written as
	\begin{equation}
		\left\{\begin{array}{l}
			x_{0}=A_{0} \sin \left(\Omega_{0} t_{0}+\varphi_0\right) \\
			v_{0}=A_{0} \Omega_{0} \cos \left(\Omega_{0} t_{0}+\varphi_0\right)\text{.}
		\end{array}\right.
		\label{19}
	\end{equation}

	Assuming at $t=t_0+\mathrm{d}t$, the oscillator obtains a stochastic impulse that changes its velocity to $v_0+\mathrm{d}v$, amplitude to $A_0+\mathrm{d}A$, phase to $\varphi_0+\mathrm{d}\varphi$, and $x_0$ unchanged. Then, we have
	\begin{equation}
		\left\{\begin{array}{l}
			x_{0}=(A_0+\mathrm{d}A) \sin \left(\Omega_{0} t_{0}+\varphi_0+\mathrm{d}\varphi\right) \\
			v_0+\mathrm{d}v=(A_0+\mathrm{d}A) \Omega_{0} \cos \left(\Omega_{0} t_{0}+\varphi_0+\mathrm{d}\varphi\right)\text{.}
		\end{array}\right.
		\label{20}
	\end{equation}
	
	Expand out the trigonometric functions in equation (\ref{20}) and substitute equation (\ref{19}) into it. We have
	\begin{equation}
		\left\{\begin{array}{l}
			x_{0}=\left(1+\frac{\mathrm{d} A}{A_{0}}\right)\left(x_{0} \cos \mathrm{d} \varphi+\frac{v_{0}}{\Omega_{0}} \sin \mathrm{d} \varphi\right) \\
			v_{0}+\mathrm{d} v=\left(1+\frac{\mathrm{d} A}{A_{0}}\right)\left(v_{0} \cos \mathrm{d} \varphi-\Omega_{0} x_{0} \sin \mathrm{d} \varphi\right)\text{.}
		\end{array}\right.
		\label{21}
	\end{equation}
	
	Make a first-order approximation to trigonometric functions in equation (\ref{21}) and ignore $\mathrm{d}A\mathrm{d}\varphi$. We can solve equation (\ref{21}) to obtain
	\begin{equation}
		\begin{aligned}
			\mathrm{d}\varphi&=-\frac{\Omega_{0} x_{0}}{v_{0}^{2}+\Omega_{0}^{2} x_{0}^{2}} \mathrm{d} v\\
			&=-\frac{1}{\Omega_{0} A_{0}} \sin \left(\Omega_{0} t_{0}+\varphi_{0}\right) \mathrm{d}v
			\text{.}
			\label{22}
		\end{aligned}
			\end{equation}
	
	As the stochastic force that induces the impulse can be written as $F_{\text {random }}(t)=\sqrt{2m\Gamma_0k_\text{B}T_0}\xi(t)$, which has been shown in section 2, and $\mathrm{d}v=F_{\text {random }}(t)\mathrm{d}t/m $, $\mathrm{d}\varphi$ can be written as
	\begin{equation}
		\mathrm{d}\varphi=-\frac{1}{\Omega_{0} A_{0}} \sqrt{\frac{2 k_{B} T \Gamma_{0}}{m}}\sin \left(\Omega_{0} t+\varphi\right)\mathrm{d}W
		\text{.}
	\end{equation}
	
	Integrate the phase varying $\Delta\varphi$ in one oscillation period $\tau_0=2\pi/\Omega_0$. We have
	\begin{equation}
		\begin{aligned}
			\Delta\varphi&=-\frac{1}{\Omega_{0} A_{0}} \sqrt{\frac{2 k_{B} T \Gamma_{0}}{m}}\int_{0}^{\tau_0}\sin \left(\Omega_{0} t+\varphi\right)\mathrm{d}W\\
			&=\frac{1}{\Omega_{0} A_{0}} \sqrt{\frac{2 k_{B} T \Gamma_{0}}{m}}\sqrt{\frac{1}{2}}W(\tau_0)
			\text{.}
			\end{aligned}
			\label{24}
	\end{equation}
	
	As the coherent time of the oscillator is much longer than $\tau_0$, using quasi-static approximations and rewriting equation (\ref{24}) into a differentiated form, we have
	\begin{equation}
		d\varphi=\frac{1}{\Omega_{0} A_{0}} \sqrt{\frac{k_{B} T \Gamma_{0}}{m}}\mathrm{d}W
		\text{.}
		\label{25}
	\end{equation}
	
	Next, we focus on the oscillator trajectory autocorrelation. For a phonon laser, ignoring the amplitude fluctuation, its trajectory can be written as
	\begin{equation}
		x(t)=A_0\cdot{\sin}(\Omega_{0}t+\varphi(t))
		\text{.}
		\label{26}
	\end{equation}
	Its autocorrelation can be written as
	\begin{equation}
		R_{xx}(\tau)=\lim\limits_{T\rightarrow\infty}\frac{1}{T}\int_{-T/2}^{T/2} x(t+\tau)x(t)\mathrm{d}t
		\text{.}
		\label{27}
	\end{equation}
	Try to solve the equation (\ref{27}). We have
	\begin{equation}
		\begin{aligned}
			R_{xx}(\tau)=&\frac{A_0^2}{2}\lim\limits_{T\rightarrow\infty}\frac{1}{T}\int_{-T/2}^{T/2} \{\mathrm{cos}[\Omega_0\tau+\varphi(t+\tau)-\varphi(t)]\mathrm{d}t-\mathrm{cos}[2\Omega_0t+\Omega_0\tau+\varphi(t+\tau)+\varphi(t)]\mathrm{d}t\}\\
			=&\frac{A_0^2}{2}\lim\limits_{T\rightarrow\infty}\frac{1}{T}\int_{-T/2}^{T/2} \mathrm{cos}[\Omega_0\tau+\varphi(t+\tau)-\varphi(t)]\mathrm{d}t\\
			=&\frac{A_0^2}{2}\mathrm{cos}(\Omega_0\tau)\lim\limits_{T\rightarrow\infty}\frac{1}{T}\int_{-T/2}^{T/2} \mathrm{cos}[\varphi(t+\tau)-\varphi(t)]\mathrm{d}t\\
			&-\frac{A_0^2}{2}\mathrm{sin}(\Omega_0\tau)\lim\limits_{T\rightarrow\infty}\frac{1}{T}\int_{-T/2}^{T/2} \mathrm{sin}[\varphi(t+\tau)-\varphi(t)]\mathrm{d}t
		\text{.}
		\end{aligned}
		\end{equation}
	
	\begin{figure}[t]
		\includegraphics[width=1\textwidth]{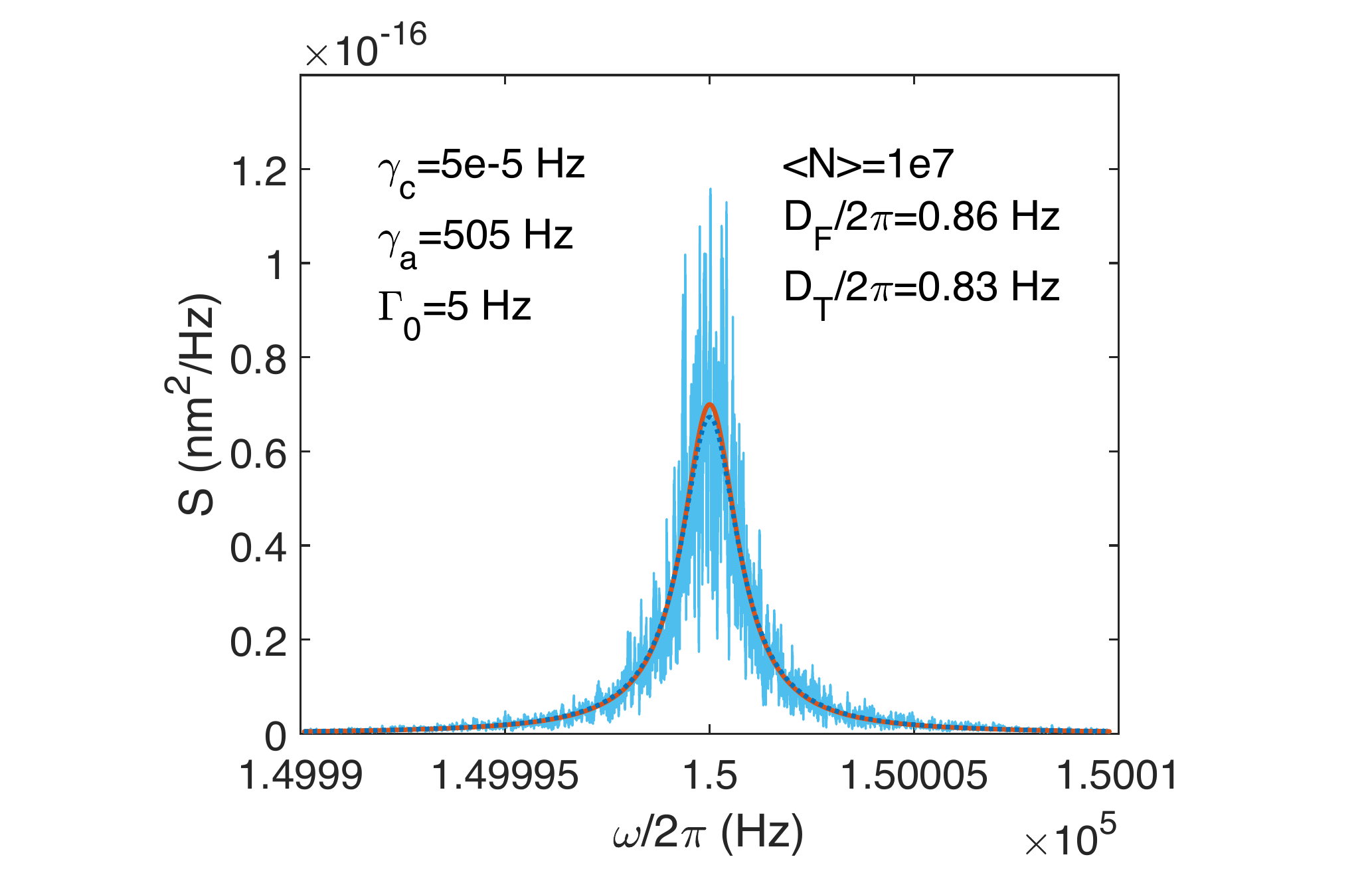}
		\caption{\textbf{Simulation of the free-run phonon laser spectrum.} The blue line is the PSD of a simulated free-run phonon trajectory. The solid line is the theoretical PSD according to equation (\ref{31}). The dashed line is a Lorentz fitting of the simulation PSD. $D_\text{T}$ is the theoretical diffusion coefficient. $D_\text{F}$ is the fitted diffusion coefficient. The simulation condition is $\gamma_c=5\times10^{-5}\text{ Hz}$, $\gamma_a=505\text{ Hz}$, $\Gamma_0=5\text{ Hz}$, and $T_0=298\text{ K}$. The length of the simulation time is 500 s.}
		\label{figS2}
	\end{figure}
	
	According to equation (\ref{25}) and the property of the Wiener process, we have
	\begin{equation}
		\begin{aligned}
			R_{xx}(\tau)
			=&\frac{A_0^2}{2}\mathrm{cos}(\Omega_0\tau)\lim\limits_{T\rightarrow\infty}\frac{1}{T}\int_{-T/2}^{T/2} \mathrm{cos}[\frac{1}{\Omega_{0} A_{0}} \sqrt{\frac{k_{B} T \Gamma_{0}}{m}}W(\tau)]\mathrm{d}t\\
			&-\frac{A_0^2}{2}\mathrm{sin}(\Omega_0\tau)\lim\limits_{T\rightarrow\infty}\frac{1}{T}\int_{-T/2}^{T/2} \mathrm{sin}[\frac{1}{\Omega_{0} A_{0}} \sqrt{\frac{k_{B} T \Gamma_{0}}{m}}W(\tau)]\mathrm{d}t\\
			&=\frac{A_0^2}{2}\mathrm{cos}(\Omega_0\tau)\mathrm{exp}(-Dt)
		\text{,}
		\end{aligned}
		\end{equation}
	where $D=\frac{k_\text{B}T_0\Gamma_0}{2m\Omega_0^2A_0^2}$ is the diffusion coefficient.
	With the mean phonon number $\left<N\right>$, we have $A_0=\sqrt{\frac{2\left<N\right>\hbar}{m\Omega_0}}$ and $D=\frac{k_{\mathrm{B}} T_{0} \Gamma_{0}} { 4\langle N\rangle \hbar \Omega_{0}}$.
	
	To obtain the power spectrum density (PSD), the Wiener–Khinchin theorem is used in the Fourier transform of the autocorrelation $R_{xx}$. We have
	\begin{equation}
		S(\omega)=\frac{A_0^2D(D^2+\Omega_0^2+\omega^2)}{(D^2+\omega^2)^2+2(D^2-\omega^2)\Omega_0^2+\Omega_0^4}
		\text{.}
		\label{30}
	\end{equation}
	Using two approximation conditions, which are $\Omega_0\gg D$ and $\omega\simeq\Omega_0$, equation (\ref{30}) can be simplified to
	\begin{equation}
		S(\omega)=\frac{A_0^2}{2}\frac{D}{D^2+(\omega-\Omega_0)^2}
		\text{.}
		\label{31}
	\end{equation}
	Equation (\ref{31}) is a Lorentzian function. It has a linewidth that reads
	\begin{equation}
		\Delta\omega_\text{FWHM}=2D=\frac{k_{\mathrm{B}} T_{0} \Gamma_{0}} { 2\langle N\rangle \hbar \Omega_{0}}
		\text{,}
		\label{32}
	\end{equation}
	or
	\begin{equation}
		\Delta f_\text{FWHM}=\frac{2D}{2\pi}=\frac{k_{\mathrm{B}} T_{0} \Gamma_{0}} { 4\pi\langle N\rangle \hbar \Omega_{0}}
		\text{.}
		\label{33}
	\end{equation}
	
	The simulation result of the free-run phonon laser spectrum is shown in Fig. \ref{figS2}. The simulated PSD well matches the theoretical PSD in equation (\ref{31}).

	\subsection{Duffing nonlinearity induced linewidth widening}
	
	\begin{figure}[b]
		\includegraphics[width=1\textwidth]{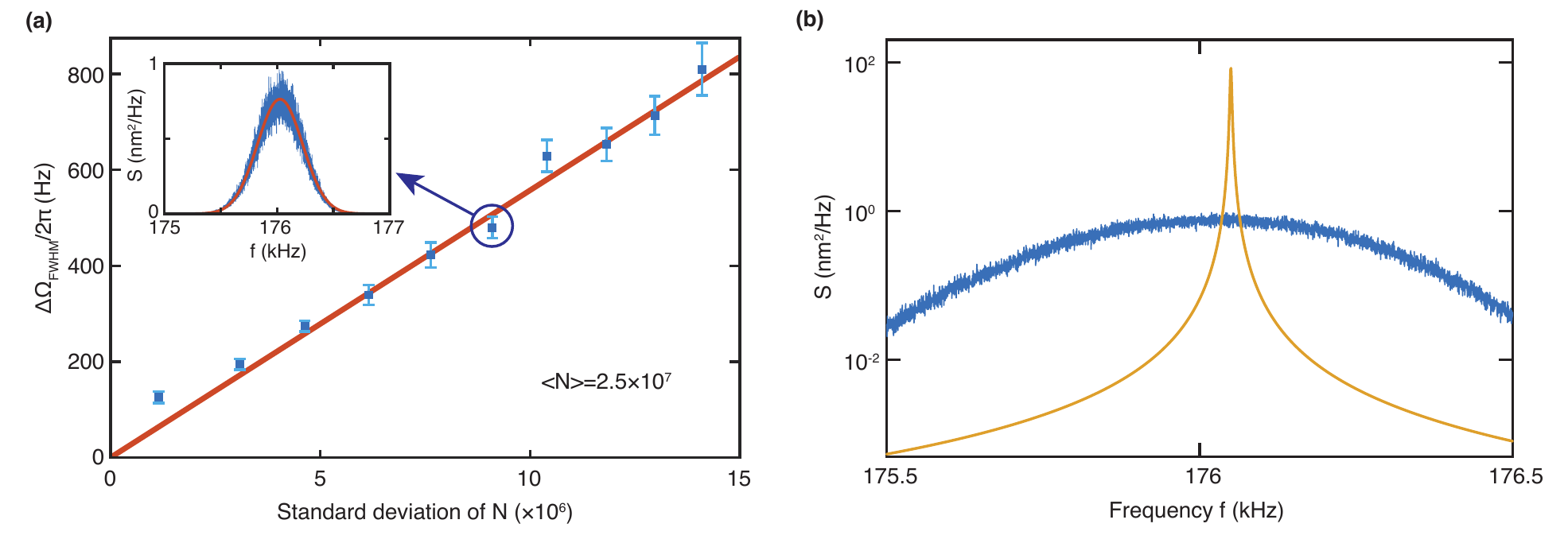}
		\caption{\textbf{Experimental verification of nonlinearity widening. a,} $\Delta\Omega_\text{FWHM}$ as a function of the standard deviation (s.d.) of phonon number without active frequency stabilization. The mean phonon number remains constant during the experiment. The solid line is a fitting of equation (\ref{37}). Error bars represent the s.d. that are calculated from 10 measurements of a 5s trajectory. The inset figure is the power spectral density of the selected data point. The smooth solid line in the inset is a fitting of the Gaussian function. \textbf{b,} Comparing the measured PSD with the theoretical PSD of a harmonic free-run phonon laser. The measured PSD (blue line) is from the insert figure in \textbf{a}. The solid yellow line is calculated from equation (\ref{31}) with the experimental conditions.}
		\label{figS3}
	\end{figure}
	
	However, due to the Duffing nonlinearity, it is difficult to experimentally obtain a free-run phonon laser with a PSD in equation (\ref{31}). As the optical potential along the trapping position is a Gaussian distribution, when the oscillator's amplitude is large, the oscillator's restoring force is in a Duffing nonlinear form, which is $F=k(x+\xi x^3)$, where $\xi=-2/w^2$ and $w$ is the $1/e^2$ beam intensity radius at the trapping point. One of the effects of the nonlinearity is that there will be a frequency shift with different oscillation amplitudes, which is $\Delta\Omega=-\frac{3}{4}A^2_0\Omega_0/w^2$. Rewriting the frequency shift in phonon number form yields
	\begin{equation}
		\Delta\Omega=-\frac{3\hbar N}{2mw^2}
		\text{.}
		\label{34}
	\end{equation}
	
	According to the phonon number distribution (equation (6) in the main text) of a phonon laser. The nonlinearity induced frequency shift also has a distribution, which is
	\begin{equation}
		\rho(\Delta\Omega) =\frac{1}{Z_\Omega} \exp \left\{-\beta_{0}\left(\frac{\hbar \Omega_{0} \gamma_{c}}{2 \Gamma_{0}}\left[-\frac{2mw^2}{3\hbar }\Delta\Omega-\frac{\left(\gamma_{a}-\Gamma_{0}\right)}{\gamma_{c}}\right]^{2}\right)\right\}
		\text{.}
		\label{35}
	\end{equation}
	
	Assume that the base frequency is a single frequency oscillation with a constant amplitude. The height of the spectrum at $\Omega_0+\Delta\Omega'$ is proportional to $\rho(\Delta\Omega')$.
	The nonlinear frequency shift induced spectrum width can be obtained by solving the two roots of $\rho(\Delta\Omega_\text{half\_max1(2)})=1/2Z_\Omega$. And $\Delta\Omega_\text{FWHM}=|\Delta\Omega_\text{half\_max1}-\Delta\Omega_\text{half\_max2}| $, which is
	
	\begin{equation}
		\Delta\Omega_\text{FWHM}=\frac{3\hbar }{mw^2}\sqrt{(\ln{2})\frac{2 \Gamma_{0}}{\beta_{0}\hbar \Omega_{0} \gamma_{c}}}
		\text{.}
		\label{36}
	\end{equation}
	Or write with the standard deviation of phonon number $\sigma_N=\sqrt{k_\text{B}T_0\Gamma_0/\hbar\Omega_0\gamma_c}$, which is
	\begin{equation}
		\Delta\Omega_\text{FWHM}=\frac{3\hbar }{mw^2}\sigma_N\sqrt{2\ln{2}}
		\text{.}
		\label{37}
	\end{equation}
	
	With the same simulation conditions shown in Fig. (\ref{figS2}) and $w=550\text{ nm}$, $m=3\times10^{-18}\text{ kg}$, we have the nonlinearity induced width is $\Delta\Omega_\text{FWHM}=2\pi\times187.9\text{Hz}$, which is much larger than the phase noise induced width of $\Delta\omega_\text{FWHM}=2D_\text{T}=2\pi\times1.6\text{Hz}$.

	The experimental verification of the Duffing nonlinearity induced width is shown in Fig. \ref{figS3}. That is, $\Delta\Omega_\text{FWHM}$ is proportional to the phonon number deviation $\sigma_N$, and the PSD can be fitted with a Gaussian function, as the phonon number distribution is also a Gaussian function.
	
	A narrower phonon number distribution can mitigate nonlinearity-induced spectrum widening. However, due to the limitation of control precision, it is difficult to eliminate the nonlinearity's influence without feedback frequency stabilization.

 %